\documentclass[12pt,journal,compsoc]{IEEEtran}

\usepackage[dvipsnames]{xcolor}
\usepackage{epsfig}
\usepackage{enumerate}
\usepackage{amssymb}
\usepackage{amsmath}
\usepackage{amsthm}
\usepackage{algorithm}
\usepackage{algorithmic}
\usepackage[english]{babel}
\usepackage{graphicx}
\usepackage{subfigure}
\usepackage{hyperref}
\usepackage{multicol}
\usepackage{xcolor}
\usepackage{soul}
\usepackage{multirow}
\usepackage{balance}
\usepackage{graphicx}
\usepackage{times}
\usepackage{xspace}
\usepackage{comment}
\usepackage{tikz}
\usepackage{pifont}
\usepackage{scrextend}
\usepackage{booktabs}
\usepackage{enumitem}
\usepackage{cite}
\usepackage{textcomp}

\usepackage{url}

\usepackage{breakurl}

\addtokomafont{labelinglabel}{\sffamily}
\definecolor{blueX}{HTML}{000699}

\renewcommand{\ttdefault}{cmtt}
\renewcommand{\encodingdefault}{OT1}
\DeclareFontFamily{\encodingdefault}{\ttdefault}{\hyphenchar\font=`\-}
\hypersetup{
	pdfpagemode=pagewidth,
	plainpages=false,
	colorlinks,
	urlcolor=blue,
	linkcolor=blue,
	citecolor=red,
	bookmarksnumbered
}

\newcommand{\eg}{{\em e.g.,}\xspace}

\newcommand{\BfPara}[1]{{\noindent\bf#1.}\xspace}

\title{Gruut: A Fully-Decentralized P2P Public Ledger}

\author{
	\IEEEauthorblockN{{DaeHun Nyang}\\
	Gruut Networks
	}
}

\IEEEtitleabstractindextext{%
\begin{abstract}
Owing to Satoshi Nakamoto's brilliant idea, a P2P public ledger is shown to be implementable in anonymous network. Any Internet user can then join the anonymous network and contribute to the P2P public ledger by providing their computing power or proof-of-work.
The proof-of-work is a clever implementation of one-CPU-one-vote by anonymous participants, and it protects the Bitcoin ledger from illegal modification. To compensate the nodes for their work, a cryptocurrency called Bitcoin is issued and given to nodes. However, the very nature of anonymity of the ledger and the cryptocurrency prevent the technology from being used in fiat money economy. Cryptocurrencies are not traceable even if they are used for money laundering or  tax evasion, and the value of cryptocurrencies is not stable but fluctuates wildly. In this white paper, we introduce Gruut, a P2P ledger to implement a universal financial platform for fiat money. For this purpose, we introduce a new consensus algorithm called `proof-of-population,' which is one instance of `proof of public collaboration.' It can be used for multiple purposes; as a P2P ledger for banks, as a powerful tool for payment, including micropayment, and as a tool for any type of financial transactions. Even better, it distributes the profit obtained from transaction fee, currently dominated by a third party, to peers that cannot be centralized. Energy requirements of Gruut are so low that it is possible to run our software on a smartphone or on a personal computer without a graphic card.
\end{abstract}

\begin{IEEEkeywords}
Blockchain, Bitcoin, Proof of Work, Consensus, Distributed ledger, Fiat Currency, Cryptocurrency, Peer-to-peer, Finance
\end{IEEEkeywords}}

\date{Jan. 2018}

\begin{document}

\maketitle

\section{Introduction}

A trusted third party has been assumed to be necessary for recording transactions in a trusted manner. In this model, the trust provided by the third party helps participants to make sure that the recordings are not manipulated. At the expense of providing the trust, the trust third party receives transaction fees from participants. It works well if participants in the network agree on the cost of running the trusted third party.
However, unforgeable online recording of transactions by peers (not by a third party) is necessary, if it is not possible technically, economically, or politically to set up a trusted third party that secures transactions. Also, even though there is a trusted third party already in place, it is more desirable to use a P2P ledger, if implementable, when the cost of maintaining the trusted third party is too high, not that it is technically or economically unavoidable, but that it makes undue profits. 
Providing a P2P ledger and eliminating the third party will have great impact on the market run by the trusted party by giving another competing technology not relying on the third party for recording transactions. 
Gruut\footnote{Gruut stands for grassroots movement in various business sectors by our fully-decentralized blockchain.} provides business entities working on real (not cryptocurrency) economy with a new business environment, where they need not rely on the trusted third party to manage transactions in a trusted way, and they will enjoy much lower transaction fees that are distributed evenly to peers running the network. 

\section{Gruut's vision}
Our vision is not just to introduce yet another cryptocurrency, but rather to create an alternative model to the business model relying on the trusted third party for real economy.
We propose Gruut blockchain not that we want to dismantle government authority, not that we distrust the current trust model provided by trusted third parties such as banks, card companies, and government authorities, but that we want to create a new business environment supported by peers for real economy, to supply an alternative having competitive edge over traditional single party models having high transaction costs, to provide go-green technology, and to set up p2p ledgers with better security. 
Pursuing Gruut's vision requires to develop a ledger technology that achieves three important goals: full economic (not political) decentralization, P2P ledger for real economy, and scalability.
\begin{enumerate}
    \item Economic decentralization: To replace ledgers managed by the third party with peer-to-peer public ledgers providing high level of trust, and to distribute the fee to peers that cannot be centralized. Anyone can join by installing GruutApp on smartphone to contribute to running Gruut blockchain. Every contributor on Gruut will have the equal chance to get rewards irrespective of their stake/computing power.
    \item Ledger for real economy: To build up a government-friendly ecosystem that is compatible with legacy/legal financial system. We are aiming at economic transparency so that Gruut should be incorporated into the legal financial platform dealing with transactions in fiat currency.
    \item Scalability: To provide a blockchain that scales out on demand and thus can process a high volume of transactions generated during online/offline payment.
\end{enumerate}

{\em Gruut project is an initiative to introduce the public
blockchain technology in real economy-based business areas while solving various
fundamental issues in current blockchain technologies. Running a blockchain
ledger as a platform for transactions in the real economy essentially limits
anonymity.}

\section{Issues with current public blockchain technology}

Our work is mainly inspired by Bitcoin~\cite{btc} which, in 2009, provided an idea that eliminates the centralized mint and decentralizes the ledger for online currency system. Coins in Bitcoin network are issued in a distributed manner by managing one global immutable ledger by peers. To make the ledger immutable, Nakamoto developed a novel voting mechanism that can resolve inconsistency in ledger by votes of anonymous peers. Because nodes on the Internet cannot be precisely distinguished, a mechanism is needed to realize one-node-one-vote. Identities such as IP address can be easily dominated. Bitcoin uses so called the proof-of-work (PoW) to implement effectively the conceptual idea of one-CPU-one-vote, but PoW requires a large amount of hash calculation. However, two important and innovative ideas -- anonymous voting and PoW (or Proof-of-Stake PoS) to implement it --  have caused two critical issues that prevent current cryptocurrencies from playing a role in real world economy: anonymity and decentralization.

\begin{itemize}
\item No decentralization: Hashing powers or stakes have been dominated by small group of people.
\item No economic transparency: All transactions are processed under random addresses of which owner is not identifiable. 
\end{itemize}

\BfPara{Economic decentralization} The top seven miner groups have more than 70\% of the world hashing power according to blockchain.info. Nowadays, cryptocurrencies are not considered to be decentralized economically as much as expected when first proposed. Crypto51~\cite{crypto51} provides the attack cost to mount 51\% attack against current cryptocurrencies, including Bitcoin and Ethereum~\cite{eth}. For example, launching at attack for one hour cost against Bitcoin costs 582,622 USD, and  364,099 USD against Ethereum. Proof-of-stake (PoS)~\cite{nxt,KRDO17,BGM14,BLMR14a,BLMR14b} also lets high stake holders control the ledger, and double-spending is reported to be possible. Delegated PoS (DPoS)~\cite{DPoS,DPoS2} in EOS elects only 21 block-producers in the whole world to maintain the blockchain and, obviously, is very far from the decentralization idea by peers. It is even more centralized than traditional banking systems considering that more than hundres of banks work collaboratively. To prevent centralization by a small group of people, one-person-one-vote is the only solution, but others are just approximation of it.

\BfPara{Economic transparency} Cryptocurrencies effectively realize online currency system as non-traceable as real currency. Since Bitcoin has been introduced, intense focus has been on cryptocurrency, and that triggered a flood of variant cryptocurrency systems. As of this writing, the number of cryptocurrencies is 1,384. Almost all cryptocurrencies using the blockchain idea, however, are non-traceable. As cash in real world is not easy to be traced even when it is used for crimes, such as money laundering and tax evasion, cryptocurrencies are not traceable except when they are exchanged with a real currency at an exchange. Modern society, however, moves to a more transparent economy owing to the growth of credit cards, check cards, and paychecks. The current cash model has been gradually diminished by those alternatives. Cryptocurrencies represented by Bitcoin, however, go toward an opposite direction that we are moving into.  One example that showed their non-transparency was that U.S. Securities and Exchange Commission decided not to accept Winklevoss Bitcoin Trust in March 2017 owing to its non-transparency of transactions~\cite{SECStatement}.

Now, we discuss cryptocurrency as a currency. The aim of Bitcoin designers is ambitious, but Bitcoin and other cryptocurrencies also fail to become a currency. People do not want to use Bitcoin to purchase goods, but for speculation, and there are many reasons to do so. One such reason is that it is separated from the real currency and it works within its own ecosystem, and the other is the high cost for mining:
\begin{itemize}
\item Closed ecosystem: Current ledger technologies rely on the closed ecosystem to keep miner's economic incentive stable. That is, cryptocurrency is for reward and transactions are only in the cryptocurrency.
\item Mining cost: Current blockchain ledgers are not good for processing small-valued transaction because of its high mining cost.
\end{itemize}

\BfPara{Closed ecosystem of cryptocurrency} In the current cryptocurrency networks, reward is given with the cryptocurrency in the network, and transactions in the ledger are done only in the same cryptocurrency. This closed ecosystem has the effect of keeping the value for attacker's incentive very stable, but on the other hand restricts the usage of the ledger only for cryptocurrency-related transactions. If a currency for transactions in the ledger is one and a cryptocurrency for reward is another, the economic incentive might not be suitable to attract miners. Let us assume that the Bitcoin ledger processes USD transactions, and rewards including coinbase and transaction fee are given in Bitcoin.  This currency separation will obviously discourage attackers from mining honestly but encourage them to misbehave (forging transaction blocks) when the value of Bitcoin plummets and the rewards get smaller than the transactions in blocks. When the value goes too high, transaction fees will be too high to treat low-valued transactions. This is why most cryptocurrencies are recursively defined within their networks. Even though it cannot separate currencies for transactions and for rewards, there must be a way to assess the value of the currency to give real economic incentives to miners. Cryptocurrency exchanges exist for this purpose, but the values of cryptocurrencies are greatly changing owing to speculation. 

\BfPara{Mining cost} On the other hand, cost itself for mining blocks (solving PoW) nowadays is too high to be used to transfer small value transactions. For example, one block mining by the very large mining pool AntMiner S9 costs 19,598.50 USD without counting the hardware cost~\cite{grisha}. Thus, it is quite obvious that we cannot use Bitcoin for processing transaction having small values.

Consequently, the closed ecosystem separated from the real-world economy and the high cost for mining make PoW/PoS-based cryptocurrencies inapplicable to real world transactions, such as buying goods. 

There are also several technical limitations in the current PoW-based cryptocurrency systems that prevent them from being used as a real currency for transactions:
\begin{itemize}
\item Energy: According to Bitcoin energy consumption index by Digiconomist, the number of U.S. households that could be powered by the energy used for Bitcoin mining is 4,049,860~\cite{digiconomist}.
\item Slow confirmation: It takes at least 60 mins for a block to be finalized theoretically, although it usually takes substantially more, owing to the limited block size and to the large number of transactions. There are faster coin networks such as Ethereum~\cite{eth}, using GHOST and having a faster block time (aiming at 12 seconds), but Ethereum still needs three or more minutes of confirmation time.
\item Scalability: The number of transactions that can be processed per second is only $4\sim20$, which is not scalable enough for the global scale of transaction networks.
\end{itemize}

Instead of using PoW, there are cryptocurrency systems that use PoS. PoS-based system where the stake holders put their stake to vote allegedly solves those problems. However, simultaneous forging of several chains is claimed to be possible, and even profitable; although there are still debates on the effectiveness of attacking PoS. The philosophy behind PoS is that a stake holder will behave owing for its own sake. DPoS (Delegated PoS) is a compromise between the fully-public ledger and the private ledger. Voting powers or block makers are limited to 21 nodes in EOS network~\cite{eos,eos_intro}. 
 
\section{Background}

\subsection{Anonymity and no-stake properties}
Two important attributes for a P2P public ledger in current cryptocurrency systems are {\em anonymity} and {\em no stake}\footnote{Here, we note that the stake in the PoS-based ledger is different from the stake in this context, but the stake in PoS is put to vote on a chain by anyone having no direct stake in the integrity of the block.}. That is, it is believed that a P2P public ledger for cryptocurrency should be run by anonymous peers having no stake in the integrity of transactions. Though the ``anonymity'' requirement provides transaction senders/recipients with transaction privacy or non-traceability, it should not be confused with node anonymity. That is, current cryptocurrency systems allow anyone to run a node without identification. There are two reasons for this: one is that there is no effective way to identify a person who runs a node on a global scale, and the other is allowing anonymity makes it easier to expand the ledger network.

The ``no stake'' requirement allows anyone to join the network to maintain a ledger, and lets anyone having no stake in the integrity of transactions to process them (or sign via PoW/PoS). If a transaction should be signed by nodes having a stake in the integrity of transactions, it should be a private ledger, where only nodes having a stake are allowed to join the network. Therefore, the ledger cannot provide the publicly-accepted trust, and actually they do not aim to provide it.  
Also, nodes in a private blockchain already have enough incentive to participate in the network even without explicit rewards (cryptocurrency), because if they do not participate they might suffer losses by the corrupted ledger, which is implicit rewards. Therefore, to run a public ledger on the Internet operable by random peers, we need the requirement of {\em no stake}. 

These two properties form the baseline for a public blockchain where anyone can join and leave the network. However, it is also necessary to have an effective way to resolve potential incoherence of the P2P ledger by anonymous peers having no stake in the integrity of transactions. This is a hard problem, but gracefully solved by PoW of Bitcoin~\cite{btc}.

\subsection{P2P ledger run by identified nodes having stake -- Private blockchain}
Hardness of voting in the anonymous network is the value of Bitcoin's PoW-based blockchain, but the very anonymity hinders application of blockchain technology to the real world financial platform, because the anonymity makes voting very inefficient. Institutes in the financial service sector, therefore, gets interested in ``private blockchain'' and ``consortium blockchain'' technology. If a distributed ledger run by identified (opposite to anonymous) nodes having conflict of interests (opposite to no stake) is needed, then the ledger falls into the category of private/consortium ledger. In the private/consortium blockchain network, it is possible for entities to run a network without any reward of ``coin'' or ``mining'', because entities having conflict of interests or having a stake in the private/consortium blockchain. The chains, therefore, can be regarded as just redundant sharing of databases for security purpose. We note that entities in a private or a consortium blockchain can authenticate each other easily, and thus can make consensus easily without PoW/PoS when discrepancy happens in the database. This ledger involves practical Byzantine fault tolerance (PBFT) protocols~\cite{PBFT,PBFT02}, and there is a huge research body in this area. Current experimental projects based on private or consortium blockchain have therefore focused on the security obtained by duplicate database management by entities having conflict of interests and the efficiency of coherence of distributed database systems. To deal with transactions among parties with conflicting interests, variants of PBFT protocols have been adopted for the private chain. Most of PBFT protocols are not scalable enough to be used on an Internet-size public chain network because current PBFT protocols support up to around 20 nodes~\cite{vukolic}.

\begin{figure*}[t]
	\centering
	\includegraphics[width=0.9\textwidth]{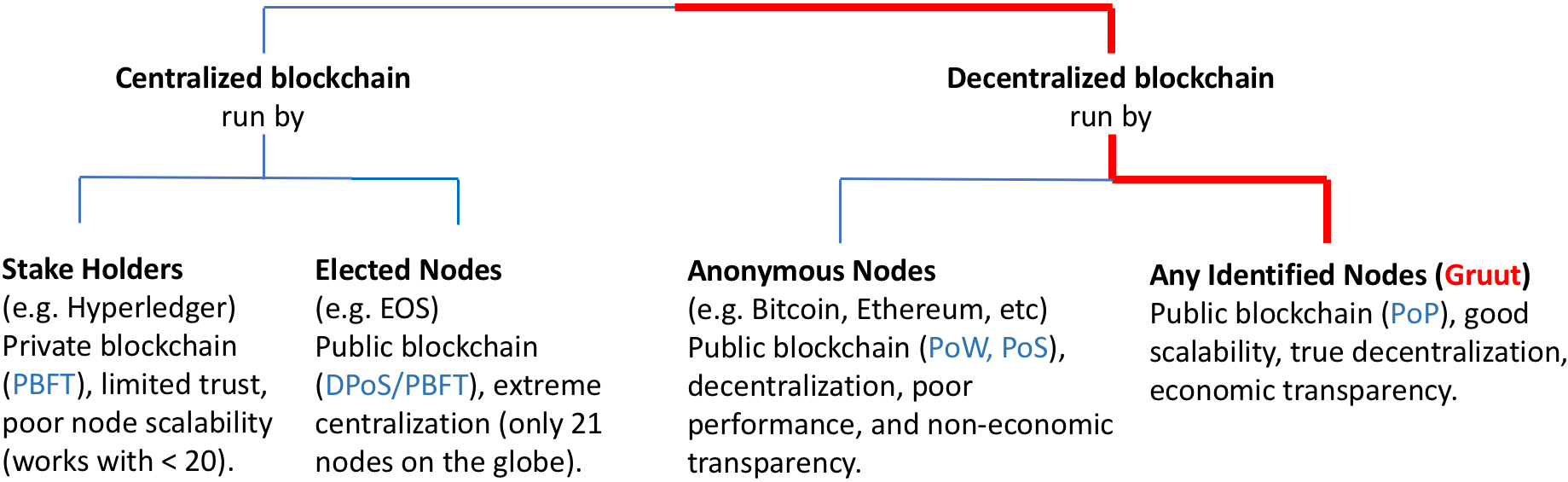}
	\caption{Design space for blockchain according to block producers and Gruut's position} 
	\label{fig-category}
\end{figure*}

\subsection{P2P ledger run by anonymous nodes having no stake -- Public blockchain}\label{sec:pub}
The problem solved by Bitcoin is sometimes called Byzantine generals problem (BGP), which is to make a consensus among generals under the assumption of faulty generals~\cite{BGP}. However, it is more exact to say that Bitcoin is an attempt to solve BGP among ``indistinguishable'' anonymous users on the Internet. Different from BGP, the problem to maintain a public ledger involves voting by anonymous users including malicious ones over the Internet, whereas works on BGP focus on how to make a consensus efficiently among already identified nodes having a stake. Bitcoin network is built to make an immutable public ledger on the Internet, where any participant with an Internet connection (and thus, obviously having no stake in the integrity of the ledger) can freely join and leave the network while remaining anonymous. 
Even worse, those participants may corrupt the public ledger for own profit. There is a high chance for multiple ledgers to coexist by corruption of peers in the network. The obvious way to make a consensus on one sole public ledger is to vote, but the voting among anonymous users in a trusted and secure manner cannot be easily done. 

Bitcoin solves the problem by PoW that effectively implements one CPU for one vote. {\em That is, PoW (similarly PoS) enforces an entity to behave as a single node instead of many Sybils. Owing to PoW, when a node behaves as a Sybil, it has a less probability to get rewards because its hashing power is divided into multiple forks (for multiple voting) and the chance to add its block decreases.} Blocks having multiple transactions together with PoW are chained to the public ledger, which makes up the blockchain. PoW is calculated by network participants called ``miners'', and the miners are rewarded with coinbase and transaction fees in cryptocurrency. To modify the ledger, an attacker forks the chain by adding a different transaction to the existing blockchain and adding PoW that at last should catch up and override the honest chain. To do so, the attacker should have more computing power to calculate PoWs for its blockchain. When it reaches the point where the manipulated blockchain is longer than the existing chain, it releases its own chain and overrides the chain for its own sake. The attacker or more possibly the group of attackers should have $51\%$ or more computing power to succeed the attack ($51\%$ attack). Cost of 50\% attack is not high in some small scale network, and cost for running 51\% attack is listed in crypto51.app~\cite{crypto51}.

\subsection{Trust by private chain vs. by public chain} 
A private/consortium ledger can be easily modified by the institute itself or by agreement of majority members of the consortium. A boss or a group of people with power can order to change the ledger. This means that only the insiders can trust the integrity of the ledger, but the ledger's integrity cannot be trusted by anyone outside the institute or the consortium. However, a public ledger should not be modifiable, and  the ledger should be able to give trust even to an outsider. That is, a public ledger must be able to provide publicly-accepted trust on the ledger's integrity, which is not the goal of consortium/private ledgers. Public trust can be acquired by letting random peers having no stake participate in the network. Gruut blockchain, therefore, should be a public chain run by random peers. 

\subsection{Sacrificing decentralization for performance}
Combining PBFT and PoW/PoS is a recent trend in designing public blockchain. Because peer-to-peer networks composed of large number of nodes are inherently slow, recent trends are to elect a small group of nodes (usually around 20 nodes) by a voting algorithm and allow only those nodes produce blocks using PBFT. Thus, researches have focused on how to enhance PBFT in terms of scalability and performance. Very recently, Thunderella has been proposed for this purpose~\cite{pass}, but this approach inevitably introduces the problem of centralization by those small number of nodes. We note that our approach is different in that Gruut does not elect any block producers but any node can be a block producer and thus, it does not raise any centralization issue.

\section{Gruut's approach and benefits}

\subsection{How to construct a fully-decentralized (or one-entity-one-vote) and low mining-cost public ledger?}\label{sec:onevote}

\BfPara{Our answer is a P2P ledger run by identified nodes having no stake} 
To build a public ledger that allows only one vote for one node is to require nodes wanting to run the network to be identified\footnote{Viable solution for online identification of a person via smartphone can be provided by the telco's identification system or by the bank's accounting system, considering that in many countries a telco and a bank must adopt the real-name accounting system called KYC (Know-Your-Customer)~\cite{kyc}. For example, in US, a Customer Identification Program (CIP) is a requirement, where financial institutions have to check the identity of individuals wanting to do financial transactions. Also, many projects on distributed KYC are ongoing these days. For example, IBM  announced proof-of-concept of its shared corporate know your customer (KYC) project with Deutsche Bank, HSBC, Mitsubishi UFJ Financial Group (MUFG) and the Treasuries of Cargill~\cite{ibmkyc}.
Also, the node can be required to connect its wallet to its bank account to join the network, which will make the online identification easy.} before joining the network. That is, Gruut does real-name (ID)-based voting (for privacy, Gruut sees only pseudonyms, so it cannot identify a node.).
Our observation is that not both anonymity and no stake attributes are needed to build a practical public blockchain, but ``no stake'' requirement is enough. Instead of requiring both attributes, we can build a public blockchain run by identified nodes requiring only no-stake property. By giving up anonymity, we can make every vote distinguishable, while allowing anyone join our network once it is identified. Therefore, one entity on Gruut has only one voting power, which enables us to implement easily double voting prevention and decentralization. 

Figure~\ref{fig-category} shows Gruut position in design space of blockchain. Gruut is so unique in the design space that it is a public ledger run by any identified nodes, whereas public ledgers are usually run by anonymous nodes and private ledgers are run by identified nodes having interest in the ledger. 

Instead of using the inefficient PoW-based voting, Gruut uses signature-based voting that has very low mining cost, 
where a node on Gruut is required only to compute a standard digital signature (such as ECDSA, RSA signature, etc). 
Computing signature for a block can be done within a millisecond even on a smartphone, and thus it lowers the entry barrier to near zero. This low mining cost contributes decentralization in terms of entry barrier. Any legal smartphone subscriber (thus, real-name identified) can join in our Gruut.

Operating environment of the public chain is different from that of the private chain. The number of nodes is huge, and the transitional forking occurs frequently owing to the large network delay. Therefore, we need a different strategy from PBFT-based majority voting algorithms taken by a private/consortium blockchain such as Hyperledger, by EOS, or by recent enhanced PBFT-based consensus algorithms~\cite{hpledger,eos,pass}. 
Our strategy is to use PoP (Proof of Population)-based majority voting in our P2P platform, which does not care who signs a block but cares how many voters sign a block. In a nutshell, PoP is a game for a miner to win rewards if the miner (we call it a ``merger'') was the first who gathers some number of signatures for a block. This is analogous to PoW, where the first miner wins the block rewards who solves the hash puzzle. 
Thus, in PoP, the number of signatures is a {\em proof} that as many {\em population} have supported the block.

The chain without anonymity requirement seems to be a private/consortium blockchain because it is run by real-name-identified nodes, but it is more of a public chain considering that blocks are signed by random nodes having no interest in the integrity of a block and that nodes can freely join and leave our network.

Benefits of using PoP as a consensus algorithm are many:
 \begin{itemize}
    \item Double voting prevention: Gruut network can prohibit one node from voting for two or more forks, because double voting is easily detectable by its identity (See section~\ref{sec:pub} for meaning of double voting prevention in PoW).
    \item Decentralization: Gruut network can prevent any single entity having large resources such as hashing power or stake (and thus, large voting powers) from centralizing the network by giving one vote for one entity, and the entry barrier to Gruut as a node is near zero.
     \item Go green: Mining cost is substantially lower than PoW-based ledger, and thus, even a smartphone can run our GruutApp as a node. 
     \item Micropayment: Owing to the low mining cost, the economic incentive for miners does not need to be very large, which means even small value transactions can be processed efficiently in the network. This implies that Gruut can be used for on/offline payment. 
 \end{itemize} 

\subsection{How to distribute rewards fairly for decentralization?}
\BfPara{Our answer is more control by the system over greed for producing blocks} 
For true decentralization, not only entry barrier should be low and one node has one vote, but also rewards should be equally distributed to peers irrespective of how much resources such as computing power or stake they have. By uniformly random algorithms, Gruut determines which nodes are to produce a block in every turn, and thus, every node has an equal chance to get rewards. Without competition, however, a node might not work hard to slow down the processing speed. To avoid this, Gruut gives a small room for competition among nodes by designating a group of nodes as candidate block-producers instead of designating only one node.  
Philosophy behind this design choice is that letting nodes compete each other without any control causes a small group of nodes who have large resources (usually, capital) to always win the race, and it eventually destroys the idea of decentralization. This is the case in Bitcoin network, Ethereum network, and EOS network. On the contrary, Gruut controls greed of nodes for producing blocks by fair algorithms, which guarantees decentralization even under any circumstances.

 \begin{itemize}
     \item True decentralization: Gruut is run by fully-decentralized peers with smartphones, and benefits are evenly distributed to peers by giving equal chances to produce blocks.
     \item Low stale block rate: Stale block is those block that are dropped owing to transitional forks. Not all the nodes but only a designated group of nodes compete to win rewards, and thus, the number of transitional forks, and thus, the stale block rate significantly decrease.
     \item Low network bandwidth consumption: Block broadcasting for gathering signatures is greatly reduced because the number of nodes competing is small. 
 \end{itemize}

\subsection{How to frustrate an attacker forking secretly for double-spending?}
\BfPara{Our answer is proof of public collaboration}
In PoP network, and to add a block, $S$ random peers among the participants make votes for the block by adding their signatures, where $S \ge 1$.  The longer a chain becomes, the more voters support the chain. Thus, when an inconsistency in the ledger occurs, a chain fork that has more supporting voters will be chosen, which is a typical implementation of majority voting.
An attacker who wants to double-spend has to be able to fork a longer chain than the main chain. 
PoP, however, is different from PoW in that an attacker cannot secretly make a dishonest fork on its own but needs a public collaboration of signers. This increases the security level drastically, because a secret forking (revealing a longer dishonest fork after secretly gathering signatures) in PoP is not possible, but a forking requires asking random signers determined by Gruut system to collude taking the risk of being reported. 


\subsection{How to deploy Gruut in mainstream business area requiring economic transparency?}
\BfPara{Our suggestion for economic transparency is identification}
In another layer, a sender and a recipient of a transaction on Gruut network may be allowed to make a transaction only after they are identified. 
This customer identification is not necessary to build Gruut network (Gruut requires node identification only), but it is for economic transparency
to avoid negative effect such as money laundering and tax evasion. Depending on the property of transactions occurring on Gruut, the level of identification may vary. For example, customer identification is mandatory for DApps dealing with financial transactions in fiat money. However, for DApp requiring censorship resistance, customer identification can be omitted. We note that Gruut blockchains are multiple, and every application domain has its own Gruut chain.
Therefore, all the transactions recorded in this network are identifiable by a government authority if needed. The new ledger can work as a universal peer-to-peer financial platform dealing with any type of asset including real currency and cryptocurrency. For privacy, pseudonyms together with public key certificates can be used. Using the platform, a peer-to-peer bank can be implemented, where the ledger of the bank is managed not in a bank's server but in a distributed immutable ledger run by peers using smart phones and computers. Roles of the bank are not to manage private ledger at its data center, but to sell financial instruments on Gruut network.

Considering that financial transactions should be conducted in real-names in almost all countries in the world, we do not need to stick to anonymous transactions. Rather, we are interested in highly-transparent public ledgers on fiat-based transactions. 
Benefit is quite obvious:
\begin{itemize}
     \item Economic transparency: High degree of transparency and traceability can be achieved, while preserving privacy (using pseudonym, among others) on the ledger. Thus, PoP transactions can work as a countermeasure on tax evasion and money laundering. 
 \end{itemize}

\subsection{How to deal with price volatility for running real financial platform?}

\BfPara{To overcome issues of closed ecosystems, our suggestion is to use a fiat currency for transaction} This design choice, and by treating transactions for legal tenders simplifies the closed ecosystem problem: the currency on our ledger is already the very currency in the real world, and thus, no concerns arise for price volatility. It also uses a cryptocurrency called ``GRU'' coin for transaction fee on the ledger. To effectively absorb the price fluctuation, all transaction fees are evaluated in terms of fiat money, and then GRU coins amounting to the price are used for the transaction fee. Therefore, GRU coin is not a security token but a utility token in that it is used as a processing fee for various types of transactions including financial transactions and content (music, movie, game, etc) purchase.
Benefits of using fiat currency are many:
 \begin{itemize}
     \item Money control by government: Fiat currency is the main currency on Gruut, and GRU coin is just an assistant medium to expedite fiat-based transactions. Governments do not need to concern themselves with ``losing control over de-facto currency''.  
     \item Immunity to price volatility: Owing to dealing with the fiat money on the ledger, it can be used freely for commerce and online/offline payment without any concern to value fluctuation.
     \item Fiat-based economy: Unlike cryptocurrency-only blockchain, Gruut is not limited to the cryptocurrency-based business area, but can penetrate into the mainstream business, owing to its power of dealing with fiat money. DApps developed and operated on Gruut can more easily deploy its business model into the real world. Combined with zero-mining cost, this will accelerate business growth.
 \end{itemize} 

\subsection{What about scalability to accommodate all the P2P transactions?}

\BfPara{Answer to the scalability issue is divide-and-conquer using multiple local blockchains in parallel and to adopt inter-chain protocol} 
All of the current blockchains try to make one global single ledger. Bitcoin has one single ledger over the entire world, and Ethereum has one too.
Keeping all transactions from all over the world in a single ledger is a suboptimal solution for scalability. 
Gruut's mining is very lightweight compared to PoW-based blockchain, which means that it is more advantageous already in terms of scalability. 
However, if the number of transactions grow, one single ledger might not be able to handle them efficiently in time. 
Gruut's approach to scalability is to introduce multiple chains, a local chain per area (e.g., one chain per country or state, or application domain). 
By doing so, the amount of transactions to be processed in one local chain will be drastically decreased, and thus the scalability issue is addressed by divide-and-conquer structure. Another interesting idea we introduce is the inter-chain transaction protocol that processes
a transaction occurring from one chain and going to another.

\subsection{What is the difference of Gruut's smart contract?}

\BfPara{Our strategy is to run a blockchain per DApp type} Heavy smart contracts slow down network throughput, and even worse they affect even non-contract transactions.
Even though everybody accepts that smart contracts are needed in many scenarios, some smart contracts require big storage and huge computation power.
We are going to divide into two types of smart contracts and run them on separate chains.
\begin{enumerate}
    \item Light smart contract: Fast and light smart contract will be included for managing ledger status (e.g., user balance). It has a deterministic Turing machine with constrained power.
\item Heavy smart contract: Customized local chains for heavy smart contracts will be used. It is optimized for running program codes, and it can produce deterministic or probabilistic results.
Probabilistic results over nodes can be merged into one result by sharing them over the block-chain. For performance, external oracle nodes (such as AWS lambdas) can be linked.
\end{enumerate}

\BfPara{Summing up, Gruut can work as a universal financial/business platform replacing traditional ones run by central trusted parties owing to an innovative consensus architecture adopting fiat-based transactions}

\section{Gruut architecture}

\begin{figure}[t]
	\centering
	\includegraphics[width=0.5\textwidth]{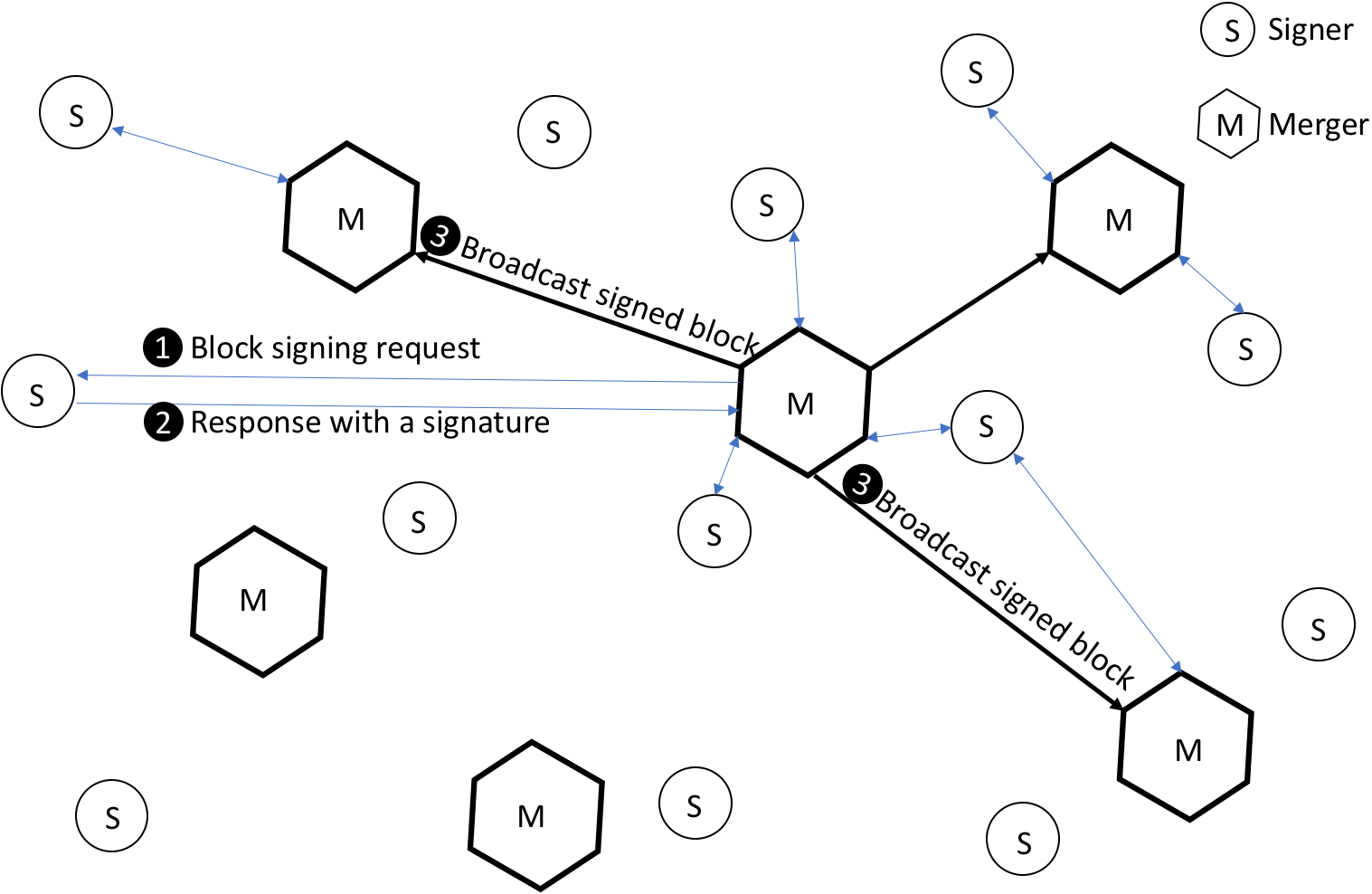}
	\caption{{\ding{202}} A merger requests signatures of a block to signers. {\ding{203}} A signer responses with a signature. {\ding{204}} Adding a block to the chain is the game that a merger wins who gathers first some pre-determined number of signatures for the block.} 
	\label{fig-network}
\end{figure}

To realize the new blockchain, PoP using digital signatures is used instead of PoW or PoS, which is a realization of explicit voting by individuals. Unlike PoW and PoS that rely on the amount of accumulated computation power or stake, PoP makes a decision according to the accumulated number of people who support a block. Thus, when a forking occurs due to an attack or mis-synchronization to produce multiple versions of the chain, the chain that has more population or the chain that is supported by more voters is chosen to be the main chain.

\subsection{Nodes: signer and merger}
There are two types of nodes in the PoP network: a node in PoP network that contributes to maintaining the public ledger by signing blocks and a node that contributes by merging transactions and block signatures into one block and chaining it to the existing blockchain. The former is called a {\em signer} and the latter node is called a {\em merger}. A signer is prompted to allow a PoP wallet to generate a signature when it gets a signing request for a transaction block. This can be automatically done by setting opt-in for the automatic signing. We expect that the signing module runs on a smart phone. By installing our signer app, a signer can make some money as a reward for signing a block. A merger collects and validates a certain number of transactions to compose a transaction block. Also, for each transaction block, it must collect some numbers of valid signatures for the block from signers and insert them into the block. This is illustrated in Figure~\ref{fig-network}. A merger's software also can run on a smart phone or on a desktop/server computer. A merger node should advertise itself or give promotions to make more signers join in their network so that it can promptly collect as many signatures as needed. When a signer joins the network, it will register itself to mergers, and a merger can ask the signer to sign a block thereafter. When a signer joins in multiple mergers, it has to set a priority among mergers.
Participating nodes in PoP network should be rewarded for their work to maintain the network. Merger nodes are rewarded for collecting/validating transactions and for collecting signatures for those transactions, and signer nodes are rewarded for their contribution by generating signatures. The reward is given to their accounts in the ledger. 

\subsection{Joining as a node on Gruut network}

To make Gruut network run, we need to make a newly-joined node identified by an authority, 
which can be seen as a bootstrapping procedure for transaction processing by peers. For the bootstrapping, we can use online identification service (or KYC) of telcos, banks, or blockchain-based distributed KYC service providers as explained in section~\ref{sec:onevote}. Especially, a node is required to connect its wallet to its bank account to withdraw, which will make the online identification easy.

\BfPara{Bootstrapping: Identification for registration} One who wants to run as a node on Gruut should be identified with a real name before entering the network. 
A user generates a public key/private key ($pub_I, prv_I$) pair and submits the public key $pub_I$ to a third party.
The third party identifies a newly-joined node and issues a pseudonym ID ($pID$) and $Cert_I$, its certificate of ($pID, pub_I$). 
It also stores ($ID$, $pID$, $Cert_I$) tuple in its database. If it is a user already registered, the registration request is rejected.

\BfPara{Pseudonym management} Gruut networks verifies the certificate $Cert_I$, 
and checks whether $pID$ is already registered or not. If everything is okay, it runs an authentication protocol to check whether the user has the corresponding private key, 
and issues a network ID ($nID$) for the pseudonym $pID$ (or a user can generate a random network ID by itself). 
Now, the user generates its own public key and the corresponding private key pair ($pub_G, prv_G$), 
and gets $Cert_G$ the public key certificate for the pair ($nID, pub_G$) from Gruut networks. 
Thus, Gruut networks' role is to issue the public key certificate to a node after confirming that it is identified by the third party. 
Gruut networks stores ($pID, nID, Cert_N$) tuple in its database.
The public key can be updated if necessary in the ledger by adding a new public key and a network ID pair. Similarly, a network ID can be updated by re-registering it in the database, but the change logs should be managed on Gruut network.

\BfPara{User} The user stores all of them including ($pID, pub_I, prv_I, nID, pub_N, prv_N$) in its wallet, but its public key and its network ID pair signed by Gruut networks is then registered in the blockchain.
A user having its certificate successfully registered in the ledger can now play a role of a signer, a merger, and/or a customer as the  user wants. 
A node can change a network ID for privacy.

\BfPara{Separation of ID registration and management for privacy} Here we note that the identification third party's role is limited to management of real-name and pseudonym pairs ($ID, pID$), 
and thus, it cannot recognize a node ($ID$) on Gruut blockchain because the identity on the chain (a signer's or a merger's identity) is $nID$ which it has no knowledge of. Also, Gruut networks cannot see who a node is because it has only a pseudonym and a network ID pair database. 
That is, the third party for identification has one half of the real name data base, and Gruut networks has the other half of it. 
Therefore, it is not possible for one entity to figure out who a node is, but two institutes must get together to reveal the identity of a node.

\BfPara{Peer-processing transactions} We also stress that neither of the identification third party nor Gruut networks involves in processing transactions, but peers in the network process them. Rewards are given to peers.

\begin{figure}[t]
	\centering
	\includegraphics[width=0.5\textwidth]{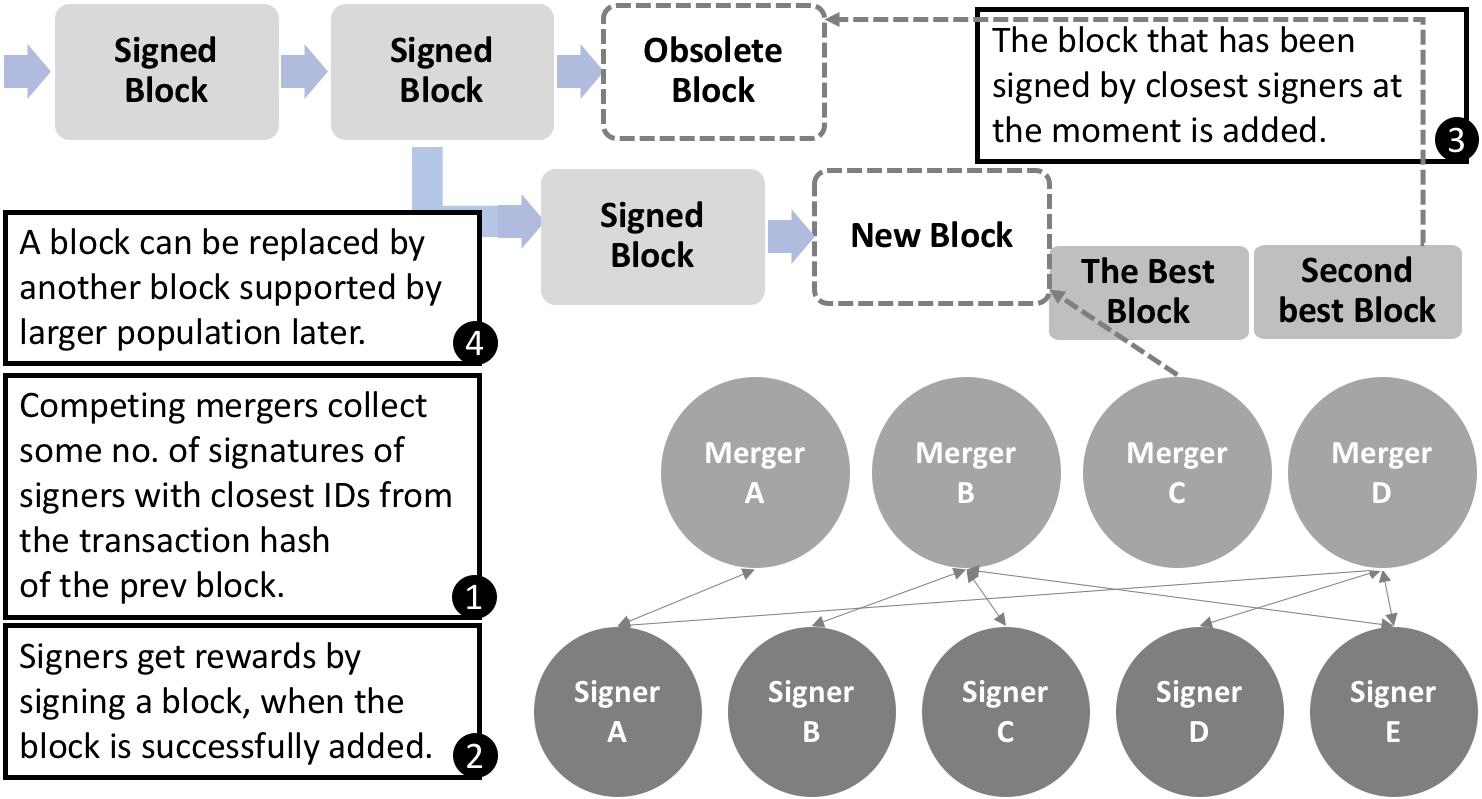}
	\caption{A block is signed by signers having closest IDs from the hash of transactions in the last block upon request of a merger. } 
	\label{fig-chain}
\end{figure}

\subsection{Processing transactions on Gruut}
When a transaction occurs from A to B, the transaction record including amount of money, date/time, transaction purpose and their IDs is signed first with A's and B's private signing keys, and then it is sent to mergers by either A or B. Mergers are chosen by the system algorithm according to a predefined hash function of the previous blocks in the chain. 
Now, a merger who gets the transaction record validates and verifies the record, and it makes a transaction block by merging some numbers of records. When a block is composed, a merger collects a predetermined (but adjusted regularly) number of signatures as soon as possible to add the block to the chain. A signer and a merger can make a signature only for a block generated after they have joined the network. Here, the signers for a block is not chosen on a merger's own, but they are chosen by a hash value, which is basically output of a random but deterministic function on the previous blocks' transaction records except signatures (similar to avoiding stake grinding attack). To change the group, therefore, it is necessary to change the previous blocks' transactions so that the hash value of them chooses the attacker's signer group. Consequently, changing a signing group of one block involves changing all the signing groups of all the preceding blocks and also of all the following blocks.

\begin{enumerate}
    \item A transaction is broadcast to the merger's network, and mergers wait until some amount of transactions are collected.
    \item A merger who has not produced any block for a period of time collects transactions into a block, computes signer ID's, finds the closest signers among the queue of signers registered on his cache roster, and sends the block to the signers.
    \item A signer checks validity of the block, and responds with its signature on the block to the merger.
    \item A merger waits until it gets some number of signatures on the block, and if successfully gathers them, it signs the block and broadcasts the multi-signed block to all mergers.
    \item Mergers accept the block only if it is valid and not double-spent, and the sending merger is eligible.
    \item Mergers notify their acceptance of the block by creating the next block in the chain, using the hash of the accepted block as the previous hash.
\end{enumerate}

A signer has to keep all the block IDs it has signed. A signer makes a signature for a block after checking 
\begin{itemize}
    \item Time stamp in the block matches with the current time considering the time synchronization error in the distributed network. If not, report to the authority and ignore the block (\eg in Bitcoin network, a block is accepted as valid if its timestamp is greater than the median of the  last $N$ blocks).
    \item The request is for a block generated before it has joined the network.
    \item The request is for a block generated before the last block it has signed and finalized. If so, the signing request from a merger will be rejected automatically. 
    \item The request is for another block having a transaction in the block that it has already signed. If so, the signer ignores it and notifies this of the merger. This notification has the effect to promptly inform the merger of the message ``you are late, and the transaction is already added.''
    \item The request is for a block including a transaction that is not consistent with one it has signed. If two blocks (regarded as a result of transitional fork) have one source transaction ID to multiple output transactions, it is regarded inconsistent. If so, the signer reports to the authority and broadcasts the merger ID and the sender ID to the network to kick him/her out of the network. All the requests will be ignored from the abuser. 
\end{itemize}

\begin{figure}[t]
	\centering
	\includegraphics[width=0.48\textwidth]{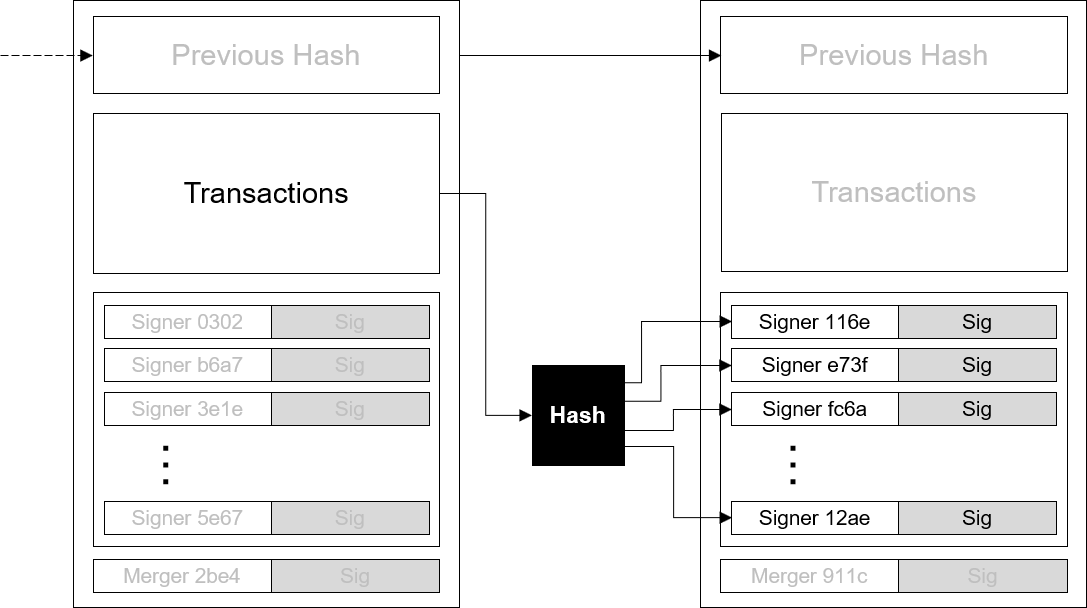}
	\caption{A block in PoP network is signed by signers and then by mergers. The signer group is determined randomly by the hash of transactions of the previous block.} 
	\label{fig-block}
\end{figure}

\subsection{Merger/Signer group selection}
\BfPara{Merger group selection} Instead of letting all mergers in the network compete, only a limited number of mergers chosen by $m_1 = H(1 || m_1' || H (T_{N-1}), T_{N-2}), \cdots, T_{N-t})$, $m_2 = H(2 || m_2' || H (T_{N-1}), T_{N-2}), \cdots, T_{N-t}), \ldots,$ $m_M = H(M || m_S' || H (T_{N-1}), T_{N-2}), \cdots, T_{N-t})$ compete among each other to get rewards, where $m_i$ is the chosen merger, $m_i'$ is the $i$-th merger in the last block, $T_i$ is all the transactions in the $i$-th block excluding signatures, $M$ is the merger group size, and $t (\ge 1)$ is the finalization threshold. This strategy can substantially reduce the number of transitional forks, and prevent random mergers from trying to forge a block by measuring merger's ID quality. 


\BfPara{Signer group selection} The signer group for a block $B_i$ is chosen according to the hash of transactions in the previous blocks (see Figure~\ref{fig-block}). That is, the list of signers is to be calculated by $s_1 = H(1 \Vert H (T_{N-1}))$, $s_2 = H(2 \Vert H (T_{N-1})), \ldots,$ $s_S = H(S \Vert H (T_{N-1}))$, where $s_i$ is the signer ID, and $S$ is the number of signatures needed to build the proof. This retrospective nature and enforced but unpredictable selection of a signing group makes corruption of the ledger much harder.

A signer/merger ID chosen by the algorithm might not be valid because it has not yet joined, or the node with the ID might not be active. 
In that case, a node having the closest ID in distance (Hamming, euclidean,  Manhattan, etc.) will be chosen. 
How close the distance is determines the quality of the merger itself and the quality of the signer group chosen by a merger. 
The distance works as a penalty for a competing chain when transitional fork occurs.

\subsection{Taking a fork supported by the largest number of voters}
When a block is added to the chain (O), it takes a few seconds to disseminate the result to the entire network to form a new chain (N). Thus, during the time, another merger may try to add another block to the chain O. This causes forking of the chain to make another chain (N') as shown in Figure~\ref{fig-chain}. The chain forking means that two versions of the ledger exist, which obviously should be resolved. On the PoP network, choosing the proper chain is done by voting. When two or more forks are competing each other to become a main chain, mergers choose the fork that has the larger accumulated number of signatures, which means that more people have supported the chain. The accumulated distance can be subtracted from the number of signers to choose the longest chain to reflect the signer group quality. The merger knows that the block possibly in the future may be dropped owing to the poor signer group quality. Thus, its strategy can be either to try to add the block with a poor signer group immediately or to search for a better signer group spending more time.  

Instead of choosing the highest score fork, we can put an additional mechanism to reduce the number of transitional forks causing unwanted transaction drop rate. That is, a merger has to wait for a time duration determined by the score of the signer group quality to add a block.

\subsection{Confirmation level} 
When a block $B_i$ is added to the chain, and the next block $B_{i+1}$ is added to that chain, $B_i$ gets one confirmation from the network. After a predetermined number of confirmations, the block is permanently added to the chain, and then another forking request will not be accepted after this permanent confirmation. Thus, the merger has to choose one among the contending forked chains that has the earliest time stamp for the last permanent confirmation. This confines the forking contention within the time frame since the last permanent confirmation. The confirmation number should be determined by the network diameter.

To enhance the throughput, we can adopt the Greedy Heaviest-Observed Sub-Tree (GHOST)~\cite{ghost,ghost_full}.
The idea is not to prune a contending branch but to leave it to support the block from which the branch has been forked.
Now, instead of choosing the longest chain, we choose the heaviest chain that the largest population has supported.

\subsection{Difficulty control: the number of signers}
New signers join in the network via mergers, and they participate in signing process from the next block. $S$ is the number of signers to sign a block. 
Difficulty level is to adjust $S$ according to the current block time statistics. It is calibrated by measuring the amount of time of 1,800 block generation, and the number is written in a block. If it is too slow to gather $S$ signatures, the minimum $S$ can be very small, for example, one.

\subsection{Scalability by local blockchains and inter-chain protocol}
For scalability of Gruut, every regional area (or application) has its own local blockchain. All transactions occurring within a region are processed as already described. So, hundreds of local chains will be running in parallel, which increases substantially the processing speed (TPS). But what if a transaction occurs across two local blockchains? When A on a local chain in New York (NYC) wants to send 5 USD to B on a local chain in Boston (BOS), "NYC sends 5 USD to B@BOS" transaction is validated first by checking balance and processed on the local chain in NYC, and then the same transaction is processed on the chain in BOS. Thus, an inter-transaction fee will be higher than an intra-transaction processing fee because it involves more mergers and signers. When B@BOS would like to send some money to C@LAX (Los Angeles), B@BOS's balance will be validated first on the chain in Boston.

Localization of chains and interoperability across chains make the ledger manageable in terms of size and speed. Because one local chain holds only the transaction records occurring in a region, the size of the ledger does not grow fast, and the transaction processing is not delayed because of the transaction volume. Also, the total TPS (transactions per second) will greatly increase because multiple local chains process transactions in parallel. Even maintaining ledgers small, transactions across chains are possible owing to the inter-chain protocol. 

PoW/PoS-based network cannot easily adopt Gruut's local chain idea because of the security problem. That is, mining is all about how much resources nodes are able to control in those networks, and thus, dispersed hashing powers (or stakes) in multiple local chains can get together to control a victim local chain with relatively small amount of resources. Gruut network runs by identified nodes, however, are not vulnerable to this attack.

\section{Miscellaneous}
\BfPara{Time scalability by hot chain and cold chain}
Owing to interoperability between chains, we can make our system time-scalable. A new chain is created, it is advertised not to use the old chain, and all the balances on the old chain are moved to the new chain (hot chain). Then, the old chain can be archived at peers' storage (full nodes) as a cold chain. If not all the balances are not transferred to the new chain and then later a transaction request on the old chain occurs, the archived chain can still be used to transfer it to the new chain.


\BfPara{Receipt on the blockchain}
Instead of storing only the money transfer information, we can store the context information also to make it possible to realize an incorruptible online receipt on the chain. To save storage space of the ledger and for better privacy, only the hash value of the transaction is stored in the ledger, while the original transaction information is separately stored by the bank. 


\BfPara{International remittance}
Obviously, it is possible to remit internationally money from an account at one chain in a country to one in another country. By applying the exchange rate, recording the transaction at the source chain first and then at the destination chain using the local chain interoperation procedure completes the international remittance.  
 

\BfPara{Digital signatures}
By Gruut signers and mergers, any secure (EUF-CMA secure) digital signature can be used in the platform, such as RSA-OAEP~\cite{OAEP}, ECDSA~\cite{ECDSA}, DSA~\cite{DSA}, Pairing-based signatures, etc. 
If the network bandwidth is not enough, an aggregate signature may be used to compress a large number of signatures~\cite{boneh}.

\section{Gruut simulation} 
This section shows simulation results conducted using a discrete event network simulator NS3 on Gruut network's performance~\cite{ns3}. 
\subsection{Simulator parameters}
\begin{table}[]
	\caption{Parameter comparison between Gruut and other blockchains}
	\centering
	\begin{tabular}{p{18mm}rrrr}
	    \toprule
		
    	& \multicolumn{1}{c}{\textbf{Bitcoin}} & \multicolumn{1}{c}{\textbf{Litecoin}} & \multicolumn{1}{c}{\textbf{Ethereum}} & \multicolumn{1}{c}{\textbf{Gruut}}\\ \midrule
		Total nodes & 10K & 2K & 16K & 10K\\ 
		\cmidrule{2-5}
		Miners & 100 & 50 & 60 & 100 mergers\\ 
		 &  &  &  & 5K signers\\ 
		 \cmidrule{2-5}
		Block generation time & 10 m & 2.5 m & 10--20 s & 10--20 s\\ 
		\cmidrule{2-5}
		Average block size & 735KB & 22KB & 25KB & 30KB\\ 
		
		\bottomrule
	\end{tabular}
	\label{Table:parameters}
\end{table}
In Table~\ref{Table:parameters}, we summarize the parameters for our simulation and compare Gruut network with other blockchain networks. In the simulation, there are 10K nodes in total, including three types of nodes: 100 merger, 5,000 signers, and 4,900 full nodes. Between nodes, we create point-to-point connection channel. The average number of connections per full node, signer, and merger are 10, 20, and 700, respectively. Assuming the use of Internet speed similar to that in South Korea, we set 39Mbps/13Mbps (download/upload) for signers and full nodes and 150Mbs/150Mbps for mergers.

We set five for the number of required signatures in each block, while the difficulty parameter is set to maintain the block interval time in the range from 15 to 20 seconds, which is used in Ethereum for reasonable finalization and security strength~\cite{12sBlockTime}. Since we focus on our PoP protocol, we did not simulate the case where transactions from normal nodes are lost due to some network conditions.

\subsection{Simulation results}
Compared to the traditional PoW protocol of Bitcoin, our PoP protocol consumes more network bandwidth by mergers to send blocks to signers to obtain signatures. In the simulation, however, only 0.054\% up to 16.25\% of transferred data is used by mergers for sending blocks and less than 0.001\% to 0.03\% of data is used by signers for sending signatures. We have conducted the simulation allowing every merger to compete to produce a block without applying merger group selection strategy. Only a limited number of mergers will compete with the strategy, and thus the bandwidth consumption and the stale block rate will be greatly reduced. 

\begin{table}[]
	\caption{Effect of the block size to other parameters. Transactions per second (TPS) is calculated with the assumption that the average size of a transaction is 0.15KB like in Ethereum. Stale block rate is the fraction of blocks discarded owing to the transitional fork. Block propagation represents the time from when a merger broadcasts a signed block to when half of nodes receive the block.}
	\centering
	\begin{tabular}{rrrrr}
	    \toprule
		
    	\multicolumn{1}{c}{\textbf{Block}} &\multicolumn{1}{c}{\textbf{Block}} &\multicolumn{1}{c}{\textbf{TPS}} & \multicolumn{1}{c}{\textbf{Stale}} &\multicolumn{1}{c}{\textbf{Block}} \\ 
    	\multicolumn{1}{c}{\textbf{Size}} &\multicolumn{1}{c}{\textbf{Interval}} &\multicolumn{1}{c}{} & \multicolumn{1}{c}{\textbf{Block Rate}} &\multicolumn{1}{c}{\textbf{Propagation}} \\ 
    	\midrule
		30.16KB & 18.56 s & 10.8 & 2.791\% & 0.78 s\\ 
		\midrule
		76.73KB & 17.58 s & 29.1 & 5.530\% & 0.99 s\\ 
		\midrule
		144.95KB & 17.60 s & 54.9 & 7.175\% & 1.56 s\\ 
		\midrule
		345.91KB & 20.42 s & 112.9 & 23.1\% & 2.91 s\\

		\bottomrule
	\end{tabular}
	\label{Table:blocksize}
\end{table}
To test the scalability of Gruut blockchain, several blockchain instances---of which block sizes are different---are examined, and the results are shown in Table~\ref{Table:blocksize}. Blocks are sent over all the nodes in the network, so it becomes harder to keep all the nodes in sync when the node size becomes bigger. As a result, mergers could fall behind other competitors in the race of creating a new block, and more forks and more stale blocks are created. This transitional forking leads to decentralization of the blockchain system including on Gruut. In the simulation with Gruut, the stale block rate is quite high when the block size reaches 300KB or when TPS reaches 100. However, considering that our PoP protocol that consumes only a small amount of network bandwidth compared to PoW protocol, which consumes a lot of energy to generate a new block, the waste of creating stale block is acceptable. Also, we note that the merger group selection strategy has not been applied to the simulation, so the stale block rate will be substantially reduced when we apply it.
 
We note that the processing speed in the simulation is just for one local chain. If $N$ local chains run in parallel over the globe, the total throughput should be one local chain's throughput multiplied by $N$. For example, we can easily get 5,500 TPS with 145KB block size (see Table~\ref{Table:blocksize}) when 100 local chains are running. 
Considering that Visa Inc. is known to process about 141.0 billion transactions in 2016~\cite{visatps} (which is 4,471 TPS on average),
our approach running multiple local chains in parallel is practical enough to cover large amount of transactions. Moreover, a number of techniques can be applied to Gruut to increase the TPS without increasing the block size. One of the most practical solutions is splitting into many transactions that can essentially be packaged into a single, smaller state on the parent blockchain, which is implemented in SegWit~\cite{segwit} and Plasma~\cite{plasma}. If both of them can add 100x to the total TPS, then Gruut can gain 10,000x in scalability.

\begin{figure}[t]
	\centering
	\includegraphics[width=0.48\textwidth]{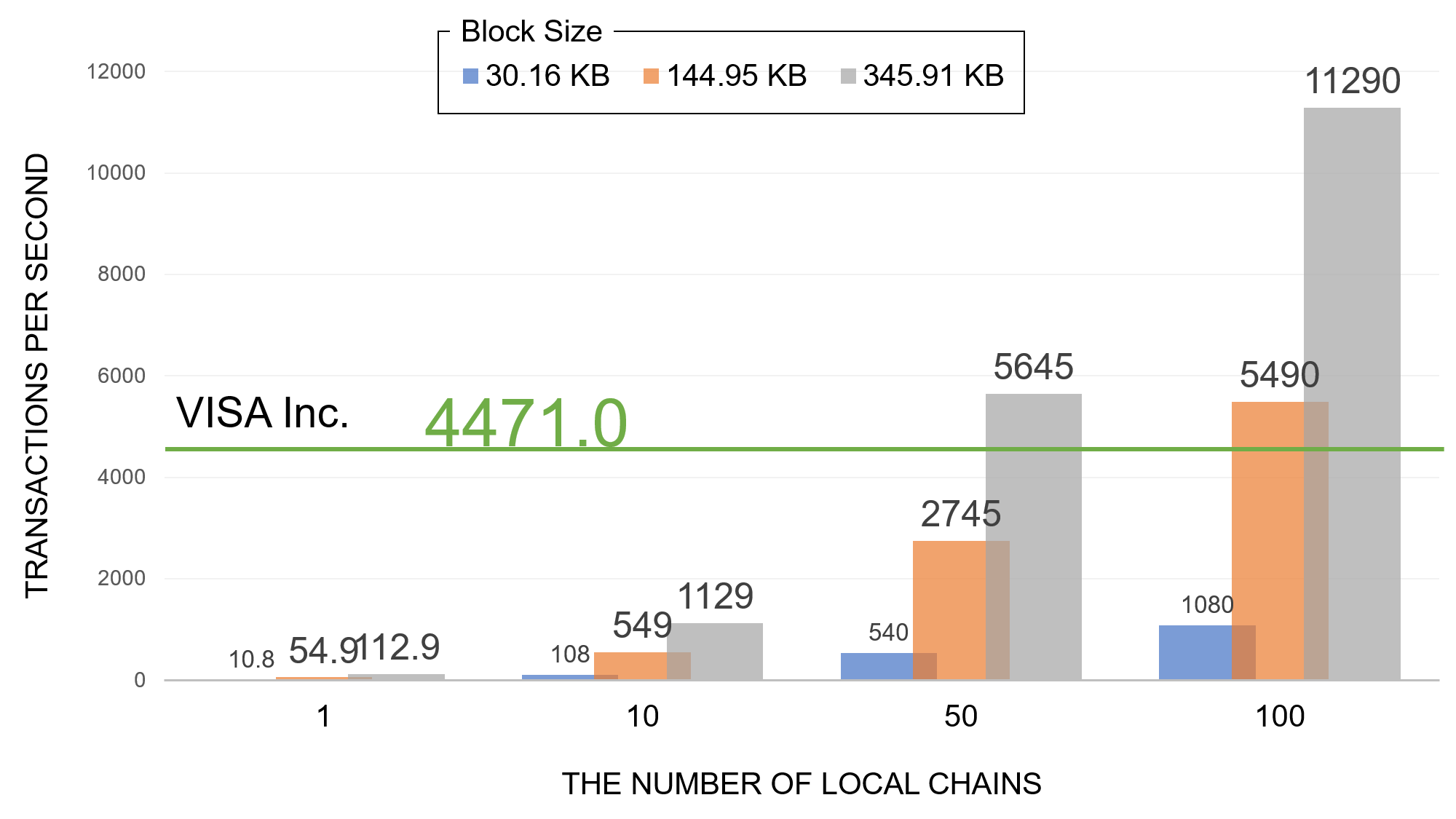}
	\caption{Extrapolation of transactions per second by the number of local chains.} 
	\label{fig-tps-chain}
\end{figure}

\section{Business development in financial sector}

Early innovations with blockchain technology have been observed in payment sector, including projects such as Bitcoin, Litecoin, Ripple, and Stellar. However, these have clear limitation of business expansion due to the fact that they use only cryptocurrency as a medium of exchange. 
As noted early, Gruut handles fiat money as its main currency and GRU coin is used in limited cases like payment of transaction fee. Of course, merchant can suggest a dual price posting where users have options to pay GRU or fiat money. Benefits of using fiat money as a main currency on Gruut are quite clear. Participants like merchant/customer and sender/recipient can enjoy volatility-resistant transactions and as a result a lot of commercial transactions will occur on Gruut Network. 

\subsection{Competitive edge over legacy financial system} 

As a single trusted party, banks and card payment player take profits for their providing trust and intermediation. By being provided with decentralized trust instead of the centralized one, business entities and customers in various business sectors will have an alternative and thus competing technology to run their business or to enjoy reduced transaction fees. Considering many intermediate institutes involve when it comes to cross-border remittance, Gruut payment system with scalable and cost-efficient technology has strong edge over legacy payment system. With this, the following fiat-based DApps can be developed: 

\BfPara{(1) P2P bank paying operational fees to peers} Any bank can run its business on Gruut ledger. They advertise their financial instruments, recruit customers by various promotions, let customers open accounts on the ledger, and manage accounts. A transaction should include a bank identifier so that multiple banks are able to run their business on the ledger. Multiple banks should compete each other on the ledger to recruit more customers or to sell more financial instruments to customers. Our P2P ledger pushes banks to make a profit by focusing on elaborating their financial instruments instead of by collecting easy money (transaction fee) automatically in exchange for using their central server system. Now, the position changes. Banks and customers have to pay the ledger fee for nodes who are running the ledger by the transaction fee. Bank can support customer's transaction fees as a promotion, and the fees are distributed fairly to nodes running the ledger. We note here that customers are opt-in to play a signer's role in the ledger network, and thus they are rewarded for processing transactions. 

\BfPara{(2) P2P credit/debit card payment} In general, card payment system fee rate ranges from 1.7\% to 3.3\% in South Korea. With the spread of Gruut payment system, it is possible to lower the payment fee rate to, say, 0.5\%. This reduction is possible because our peer-to-peer ledger does not need to spend on operational cost whereas the traditional card system has to run its own data center. A bank runs the credit card business on Gruut ledger. Here, transactions are recorded whenever a payment occurs at a merchant. Now, what a merchant should pay for the system can be greatly reduced by reduced merchant account fee, by excluding a card processor, and by not using the bank's central system. Essentially, all the transaction-related fees (substantially reduced amount) are distributed fairly to nodes supporting the ledger network.

\subsection{Two strategies to deal with fiat money}
Theoretically, money is the claim right for real values of product and service, and this claim is effective with government guarantee in the modern society. Gruut ledger running on fiat money systems is compatible with legacy financial systems, and it might be able to provide customers with the claim right. 
Simply recording fiat money transactions on Gruut ledger, however, does not make the real value transfer occur unless the belief that a transaction on Gruut ledger is a legally-binding contract is acquired.
Cryptocurrency economy is not linked with real economy (linked only at an exchange), so the matter in cryptocurrency system boils down to the belief among participants within the ecosystem. However, Gruut ledger connects and binds fiat currency with real economy, and thus we have to spread and disseminate the belief that a transaction on Gruut ledger is safely bound with the corresponding value transfer in real economy.   
To resolve the issue, we propose two strategies: one is to establish an operating company (say GruutCo.) to provide customers with payment guarantee, and the other is to let traditional banks run Gruut ledger to replace their ledgers in data center with (p2p ledger-as-a-service model) and to make them pay the fee. 

\BfPara{(1) Payment guarantee within Gruut network} A customer is required to deposit some amount of fiat money to GruutCo.\ when the user signs up, and GruutCo.\ sets the account's balance on Gruut ledger. Now, the money deposited can be freely spent  (in fiat) for payment at a merchant that accepts our Gruut payment. Later, when a merchant wants to withdraw its fiat money, GruutCo.\ can transfer money in fiat currency from its bank account to the merchant's bank account. GruutCo.\ works only as an entity providing payment guarantee, but it does not manage the ledger for transactions and it does not get any processing fee. 

\BfPara{(2) Legal contract ledger-as-a-service} Gruut ledger can be provided as a infrastructure for banks. By doing so, Gruut ledger can be regarded as one of nodes in a bank's data center, and it can acquire the status of ledger having legal contract. To make this happen, strategic alliance with current financial service platforms are necessary.
Alliance with more banks or other financial institutions means that Gruut has acquired public trust on payment guarantee by customers. On the one hand, this is a big challenge to business expansion of Gruut payment network, but on the other hand, it could be a great opportunity if we can provide a reasonable mutual benefit both to Gruut and to legacy financial institutes.
Now, most of the financial institutions keep eyes on the evolution of blockchain technology to predict the payment market trends. 
Some of them are examining how to utilize this technology to compete with other players and to survive under the new paradigm shift. 
First benefit that financial institution can obtain from Gruut is cost reduction for payment infrastructure. Especially, in cross-border remittance and card payment with payment gateway (PG) and acquirer, Gruut network is able to provide cost-effective payment infrastructure to existing financial institutions by removing and shortening complicated payment steps. In addition to cost reduction, Gruut is considerably superior to legacy financial one in terms of security. Other than diversifying hackers' attack points, Gruut provides more advanced security by using identification-based voting to enforce nodes to behave, unlike  in other blockchain networks. Benefits from cost and security aspects let existing financial institutions have enough motivation to collaborate with Gruut network. 

\subsection{Local mining and Gruut business deployment}
Unlike current blockchain projects, Gruut makes use of node identification and local chains. This limits a node to work or mine on a local Gruut ledger in its country of residence. This limitation is fundamentally different from the case of the blockchains out there, where any node can contribute to the global chain irrespective of its residency. Gruut business strategy, therefore, is not to make one global single ledger all at once, but to launch local Gruut ledgers incrementally over regions.

\section{Conclusion}
In this white paper, we have proposed a new blockchain technology called Gruut. 
By introducing node identification, customer identification, and fiat money as a main currency, the new ledger has overcome many issues that could not be solved easily by current blockchain technologies, including decentralization, incompatibility with legacy financial platform, and scalability. 
We believe that Gruut will give a new direction to blockchain technology, and that Gruut will give great impact on the current business model relying on a trusted third party gathering transaction fees by lowering the fee and distributing it to peers. 

\ \\

\BfPara{Acknowledgement} I would like to express the deepest appreciation to Brian Oh, SungWone Choi, Tom Lim, Jeonil Kang, DongOh Shin, DinhNguyen Dao, and Aziz Mohaisen for their support and valuable comments.

\clearpage
\bibliographystyle{unsrt}
\bibliography{references}

\begin{IEEEbiography}[{\includegraphics[width=1in,height=1.25in]{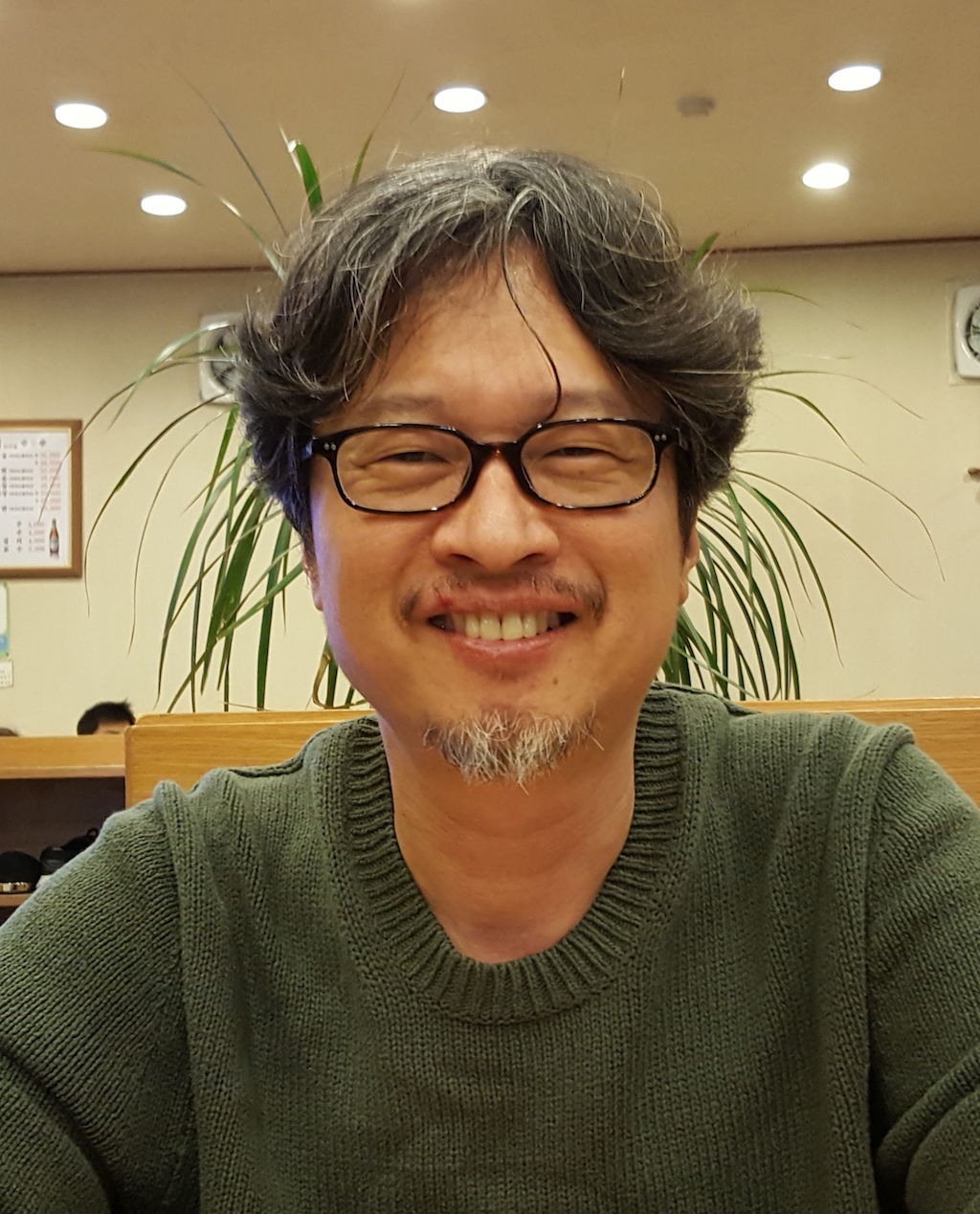}}]{DaeHun Nyang}
received double BEng degrees in electronic engineering and computer science from KAIST, and MS and PhD degrees in computer science from Yonsei University, Korea, in 1994, 1996, and 2000, respectively.
He has been a senior researcher at Electronics and Telecommunications Research Institute, Korea, from 2000 to 2003.
Since 2003, he has been a full professor in Computer Engineering Department at Inha University, Korea, where he is also the founding director of the Information Security Research Laboratory, and a CEO of theVaulters, Inc.. He also established Gruut networks to develop Gruut blockchain in 2018.
He is a member of the IEEE and a member of the board of directors and editorial board of the Korean Institute of Information Security and Cryptology.
His research interests include cryptography, network security, privacy, blockchain, cryptocurrency, traffic measurement, network monitoring, deep learning-based software code analysis, usable security, biometrics and their applications to authentication and public key cryptography.
\end{IEEEbiography}
\clearpage

\begin{appendices}
\section{}
\subsection{Forging a block}
To forge a block, an attacker has two options: one is to forge signatures in the block, which is infeasible owing to the EUF-CMA security property of the signatures. Instead of breaking the cryptographic signature, it can bribe the signing group of the block. The other way is to change the singer group and put signatures of its own signer group.

\BfPara{Replacing signer group} When an attacker modifies a transaction in one block $B_i$ to get $B_i'$, it has to get $B_i'$ signed by the signer group. If not to corrupt them, it should change the signer group. For this, the previous block $B_{i-1}$'s transaction records must be modified to $B_{i-1}'$, because the signer group is determined by $B_{i-1}$'s records. This in turn involves re-signing of the modified block $B_{i-1}'$. The signer group for the block $B_{i-1}$ won't re-sign $B_{i-1}'$ because they already signed in the time frame, and thus, the signer group for $B_{i-1}'$ should be changed. Consequently, modification of a block $B_i$ involves recursive modification of $B_{i-1}, B_{i-2}, \ldots, B_1$ in backward. Also, by changing the transaction in $B_i$, both the chain hash stored in $B_{i+1}$ and the signer group of $B_{i+1}$ are changed. Consequently, it involves both forward and backward chain modification to change one block.

\begin{figure*}[t]
	\centering
	\includegraphics[width=0.9\textwidth]{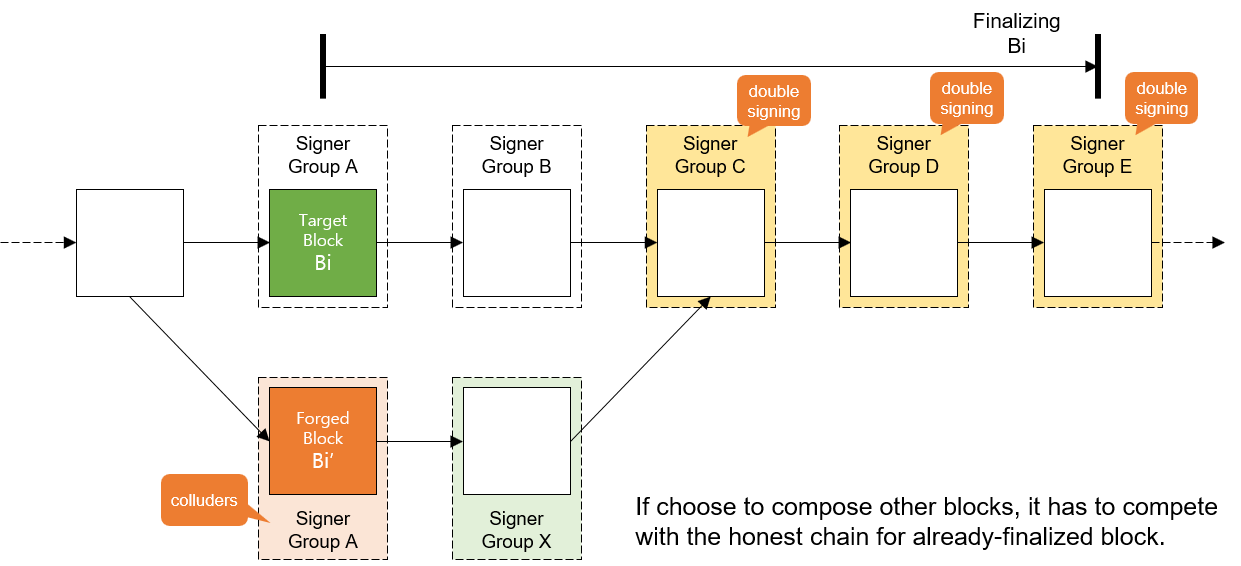}
	\caption{Forging a block is discouraged because a forger has to publicly advertise to gather colluders, and the share is less than the fee, bounty, and penalty.}
	\label{fig-forging}
\end{figure*}

\BfPara{Bribery-corrupted or stolen key-compromised signer group} By bribing a signer group or stealing a key to forge a block, an attacker does not need to change the previous blocks, because the signer group for the current block $B_i$ has not been changed, so valid signatures by the correct but corrupted signer group are added (See Figure~\ref{fig-forging}). The attacker cares only about the forward forgery. Owing to the change of transactions and corresponding signatures, the chain hash value and the signer group for the next block $B_{i+1}$ essentially does not match with those of the block in the honest chain. Therefore, the attacker must bribe or compromise again the new signer group to sign the block $B_{i+1}$. Now, the modified block $B_{i+1}'$ has a different signer group, a different chain hash, but the same transactions, which causes a different signature. For $B_{i+2}$, the very signer group that has signed the block should be corrupted by the attacker, which repeats to the last block of the chain. The required number of corrupted signers to revert the finalized voting is too large to be compensated by the amount for the forged transaction.
 
\BfPara{51\% attack} Assume that a double-spending attacker is able to control $q$ fraction of the total signers, and an honest node can $p$. Because the signer group is chosen uniformly at random, a double-spending attacker has the success probability $q$ to make the next block, and an honest node has $p$. The analysis to catch up with the honest chain by the double-spending attacker starting $z$ blocks behind is the same as that in Bitcoin network~\cite{btc}. If 51\% ($p<q$) of signers are controlled by the attacker, it is possible to catch up the honest chain. However, to be successful in this attack, the attacker should be able to bribe 51\% of signers taking the risk of being reported. 

\subsection{Security economy, bounty, and penalty}
There are multiple motivations that discourage the attack against the ledger. 

\BfPara{First and foremost, it is a fully-traceable real-name-based ledger} The strongest drive comes from the fact that every node has been already real-name identified and every transaction on the ledger is fully traceable. Thus, nodes in the network are fully aware that crimes are perfectly traceable and it is highly likely to be tracked down.

\BfPara{Second, secret fork is not possible, but public advertising for forking is necessary} In PoW network, a miner can secretly build up a fork for double spending by computing PoW faster than other miners until it catches up the honest chain, and then it releases the secret fork to replace the honest chain. However, in PoP network, a merger to forge a block cannot do this secret forking, because making a forking chain cannot but involve a large numbers of signers to collude. The signers proposed by the forging merger, however, already know that the block has been already finalized with a high probability (a block is finalized after the time of the network diameter). Thus, the attacker should take the risk that any signer will report the double spending attempt to the authority. 

\BfPara{Third, the incentive given to each colluding signer is smaller than the transaction fee} 
Let $f$ be the transaction fee fraction, and $\mathsf{TA}$ be the maximum amount limit per block. Also, a single transaction is limited less than $e$ of $\mathsf{TA}$. Let $\mathbb{S}_i$ and $R$ be the signer group  for a block and the number of blocks for finalization, respectively. The number of signers, $\vert \mathbb{S}_i \vert$, should be determined by the network speed, and $R$ by the network diameter. For simplicity, we let $\Tilde{S}$ be the average number of signers for a block.
The best case scenario for the attacker is to forge the block that has just been finalized because it needs the minimum number of blocks to be re-signed.
The attacker is forging a block so that its own transaction for spending money shall be removed. The forged block loses $e \times \mathsf{TA}$ transactions for attacker's double spending, and thus, $e \cdot f \times \mathsf{TA}$ should be compensated to the colluding signers of the forged block. The signer group for the next block now changes to a new one, and they have to sign the next block to the forged one. The following blocks are signed by the same signer groups because the transactions do not change, so they don't get any extra income from the new signing. Here, the colluding signers for the following blocks are asked to sign again the same block while they know it is double signing act. Now, they have to choose whether to get in the collusion or not. They have two concerning points: one is that by network nodes, it is highly likely to be reported for the double signing act. The other is that they know the double signing act might lose their chance to get the transaction fee. This is because the former chain they have signed has obviously better chance to win the transaction fee than the later chain that they are about to collude has. To sum up, the share a colluding signer can get is
\begin{align}
    \frac{e \times \mathsf{TA} - e \cdot f \times \mathsf{TA} }{R\times \Tilde{S}} \label{eqn:atkbft}
\end{align}
Transaction fee that can be achieved by behaving well is 
\begin{align}
    \frac{ f \times \mathsf{TA} }{\Tilde{S}} \label{eqn:normbft}
\end{align}
Therefore, a signer will compare Eq.(\ref{eqn:normbft}) to Eq.(\ref{eqn:atkbft}). By simple algebra, we can get $f^{\prime} = e/(R+e)$. When $f$ is greater than or equal to $f^{\prime}$, a signer will behave. If we set $e=0.1(10\%)$, and $R = 10$, then we can set $f^{\prime}$ be less than $0.01(1\%)$. This means that the transaction fee $1\%$ is enough to prevent colluding attack if the maximum portion of a single transaction in a block is limited to $10\%$ and a block is confirmed after 10 blocks including the block are added.

\BfPara{Fourth, bounty and penalty} To protect the network, we can introduce the bounty given to the reporter of double spender and the penalty to the attacker.
A potential attacker is not likely to try to forge the ledger if the penalty is more than the profit by the forgery considering that it should publicly advertise and recruit the signers to collude. 
A random signer is more prone to report the double spender because the bounty is set to be higher than the share it can get when it colludes for illegal forgery.
Both the bounty and the penalty should be set to be higher than $((1 - f)\cdot e\times \mathsf{TA} )/(R\times \Tilde{S})$. 

\BfPara{Last, the incentive to be given by the attack is limited by the system policy that sets the maximum transaction value (\eg $10\%$ of the total in a block)} 
If a transaction is high in price, the transaction should be split into many small transactions to discourage the attack. Max profit that colluders get is limited. By setting $\mathsf{TA}$ small for each block, the incentive of attackers gets small. For example, if we set $\mathsf{TA}$ USD 10,000 and $\Tilde{S}$ $10$, the maximum profit for each colluder is only USD 9.9 considering that $R=10$, $e=10\%$, and $f=1\%$. USD 9.9 does not seem to be enough to take part in the crime in a fully traceable network. 

\end{appendices}

\end{document}